\documentclass{amsart}

\usepackage[dvips]{graphicx}

\newcommand{\R}{\mathbb R}

\newcommand{\T}{\mathcal T}

\def\b{\begin{eqnarray}}
\def\e{\end{eqnarray}}
\def\v{\varphi}
\def\n{\noindent}

\def\d{\delta}
\def\i{\int\limits_{0}^{L}}
\def\j{\int\limits_{0}^{\eta}}

\begin{document}

\title{Nearly-Hamiltonian structure for water waves with constant vorticity}%

\author{Adrian Constantin, Rossen I. Ivanov and Emil M. Prodanov}%

\address{Trinity College, School of Mathematics, Dublin 2, Ireland}
\email{adrian@maths.tcd.ie}
\address{Trinity College, School of Mathematics, Dublin 2, Ireland}%
\email{ivanovr@maths.tcd.ie}%
\address{Trinity College, School of Mathematics, Dublin 2, Ireland}%
\email{prodanov@maths.tcd.ie}%

\maketitle


\begin{abstract}
We show that the governing equations for two-dimensional gravity water waves 
with constant non-zero vorticity 
have a nearly-Hamiltonian structure, which becomes Hamiltonian for steady waves. 
\end{abstract}



\section{Introduction}

The mathematical study of water waves was initiated within the framework of 
linear theory with the work of Airy, Stokes, and their contemporaries in 
the nineteenth century. Periodic two-dimensional water waves are of special 
interest since the typical water waves propagating on the surface of the sea 
(or on a river or lake) present these features.  Stokes \cite{St} noticed that actual water wave 
characteristics deviate significantly from the predictions of linear theory. This 
started an extensive study of the nonlinear governing equations for water waves.

A celebrated development in water-wave theory was the discovery by Zakharov \cite{Z} 
that the governing equations for two-dimensional irrotational gravity water waves 
have a Hamiltonian structure - see the discussions in \cite{BO, C, CG}. The aim of 
this paper is to present a nearly-Hamiltonian formulation for two-dimensional gravity 
water waves with constant vorticity. For irrotational flows (zero vorticity) we recover 
Zakharov's result. Moreover, if we restrict our attention to steady waves, a Hamiltonian 
structure emerges; we refer to \cite{GT} for an in-depth discussion of the Hamiltonian structure 
of steady irrotational water waves. While Zakharov's ideas were generalized in various 
directions by several authors - see \cite{Co0, KS} for a survey of contributions in this 
direction, the elegance and simplicity of the nearly-Hamiltonian formulation for 
flows of constant vorticity makes it conceivable that it might be instrumental in 
deriving qualitative results for such flows. Related to this possibility, recently \cite{CS1} 
stability results for steady water waves with vorticities that depend montonically on 
the depth were derived from the variational formulation provided in \cite{CSS}. The 
reason for considering water waves with constant vorticity is twofold. Firstly, by the 
Kelvin circulation theorem \cite{J, L} a two-dimensional water flow that is initially 
of constant vorticity remains so at later times and therefore the restriction to such flows 
is justifiable. Secondly, as already pointed out, the elegance of the nearly-Hamiltonian 
formulation valid within this setting is mathematically attractive. From the physical 
point of view we notice that while 
irrotational flows are appropriate for waves propagating into a region of still water \cite{L}, 
water waves with vorticity describe wave-current interactions - see the discussions in 
\cite{Co, SCJ, Th}. Tidal flows are the most prominent example of water flows with constant 
vorticity \cite{DP}.  

\vfill\eject

\section{Preliminaries}

To describe two-dimensional periodic water waves it suffices to consider a cross section of the flow 
that is perpendicular to the crest line. Choose Cartesian coordinates $(x,y)$ with the 
$y$-axis pointing vertically upwards and the $x$-axis being the direction of wave 
propagation. Let $(u(t,x,y),\,v(t,x,y))$ be the 
velocity field of the flow, let $\{y=0\}$ be the flat bed, 
and let $\{y=\eta(t,x)\}$ be the water's free surface. 

For gravity water waves the restoring force acting on 
the water's free surface is gravity and the effects of surface tension are neglected. Assuming the 
water density to be constant ($\rho=1$) - this is physically reasonable cf. \cite{L}, we obtain the 
equation of mass conservation 
\begin{equation}
u_x+v_y=0.
\end{equation}
Appropriate for gravity waves is also the assumption of inviscid flow \cite{L}, so that 
the equation of motion is Euler's equation
\begin{equation}
\begin{cases}
u_t+uu_x+vu_y = -P_x, \\
v_t+uv_x+vv_y = -P_y -g,
\end{cases}
\end{equation}
where $P(t,x,y)$ is the pressure and $g$ is the gravitational constant of 
acceleration. The free surface decouples the motion of the water from 
that of the air so that 
\begin{equation}
P=P_{atm}\qquad \hbox{on}\quad y=\eta(t,x),
\end{equation}
must hold, where $P_{atm}$ is the atmospheric pressure \cite{J}. Since the 
same particles always form the free surface, we have 
\begin{equation}
v=\eta_t+u\eta_x\qquad\hbox{on}\quad y=\eta(t,x).
\end{equation}
On the flat bed the boundary condition
\begin{equation}
v=0\qquad\hbox{on}\quad y=0,
\end{equation}
expresses the fact that water cannot permeate the rigid bed $y=0$. 
The governing equations for periodic two-dimensional gravity water waves propagating over a flat bed are 
(1)-(5), with the specification that the periodicity is reflected in the fact that all functions 
$u,\, v,\,P,\,\eta$ exhibit a periodic dependence in the $x$-variable of, say, period $L>0$. 
Other than the nonlinear character of the equations, the main difficulty in their analysis 
lies in the fact that we deal with a free-boundary value problem: the free surface $y=\eta(t,x)$ is 
not known {\it a priori}. Throughout this paper we consider flows of constant vorticity, that is, 
throughout the fluid the vorticity
\begin{equation}
\omega=v_x-u_y
\end{equation}
is constant ($\omega \in \R$) throughout the fluid domain
$$\Omega(t)=\{(x,y) \in \R^2:\ 0<x<L,\ 0<y<\eta(t,x)\},$$
the free surface of which is given by the graph
$$S(t)=\{(x,y) \in \R^2:\ 0<x<L,\ y=\eta(t,x)\}$$
restricted to a period cell. Furthermore, we require that
\begin{equation}
\int_0^L u(t,x,0)\,dx=0,\qquad t \ge 0.
\end{equation}
The relevance of this last condition is explained in Section 3.

\section{The nearly-Hamiltonian formulation}

For two-dimensional flows the incompressibility condition (1) ensures the existence of 
a stream function $\psi(t,x,y)$ determined up to an additive term that depends solely 
on time by
\begin{equation}
u=\psi_y,\quad v=-\psi_x.
\end{equation}
Since (5) becomes $\psi_x(t,x,0)=0$, we can determine $\psi$ uniquely by setting 
$\psi=0$ on the flat bed $y=0$, that is, we set
\begin{equation}
\psi(t,x_0,y_0)=\int_{0}^{y_0} u(t,x_0,y)\,dy\quad\hbox{for}\quad (x_0,y_0) \in \Omega(t).
\end{equation}
This explicit formula shows that $\psi$ is $x$-periodic with period $L$. 
In terms of the stream function, the vorticity $\omega$ is determined from (6) by
\begin {equation}
\Delta \psi=-\omega\quad\hbox{in}\quad \Omega(t).
\end{equation}
Let us now introduce the (generalized) velocity potential $\v(t,x,y)$ via
\begin{equation}
u=\v_x-\omega y,\quad v=\v_y.
\end{equation}
Notice that this is not the standard reduction by the Weyl-Hodge decomposition (see 
\cite{CDG, W0}) since the vector field $W=(-\omega y,\,0)$ is divergence free but does not 
satisfy the boundary condition $W\cdot n=0$ on a free surface that is not flat ($n$ 
being the outward normal to the boundary). The value of the (generalized) 
potential $\v$ at $(t,x_0,y_0)$ can be determined by integrating $\v_x$ along the 
horizontal segment with endpoints $(0,0)$ and $(x_0,0)$, and subsequently $\v_y$ along the 
vertical segment joining $(x_0,0)$ to $(x_0,y_0) \in \Omega(t)$:
\begin{eqnarray*}
\v(t,x_0,y_0) &=& \v(t,0,0)+\int_0^{x_0}\v_x(t,x,0)\,dx+\int_{0}^{y_0} \v_y(t,x_0,y)\,dy \\ 
&=& \v(t,0,0)+\int_0^{x_0}u(t,x,0)\,dx+\int_{0}^{y_0} v(t,x_0,y)\,dy.
\end{eqnarray*} 
The potential $\v$ is a harmonic function since
\begin{equation}
\Delta\v=0\quad\hbox{in}\quad \Omega(t).
\end{equation}
Notice that, independently of the additive time-dependent term up to which $\v$ is uniquely 
determined by (11), we have by the above explicit formula for $\v$ that
$$\v(t,x_0+L,y_0)-\v(t,x_0,y_0)=\int_{x_0}^{x_0+L} u(t,x,0)\,dx,
\qquad (x_0,y_0) \in \Omega(t).$$
In view of the $x$-periodicity of the function $u$, we see that the right-hand side equals 
$\displaystyle\int_0^L u(t,x,0)\,dx$. At this point some results 
on steady water waves (water waves for which the free surface $\eta$, the pressure 
$P$ and the velocity field $(u,v)$ exhibit an $(x,t)$-dependence of the form $x-ct$, where 
$c \neq 0$ is the wave speed - that is, in a frame moving at speed $c$ these waves are stationary) 
are of relevance. In the irrotational case, the existence of steady waves of large amplitude 
(that is, waves that are not small perturbations of a flat surface) satisfying 
$\displaystyle\int_0^L u(t,x,0)\,dx=0$ was established in \cite{AT} - see also the discussions 
in \cite{Co2, T}. For these waves, called {\it Stokes waves}, we therefore have that the 
velocity potential $\v$ is $L$-periodic in the $x$-variable. On the other hand, for flows of 
constant vorticity $\omega \in \R$, there are steady waves of large amplitude for which 
$\displaystyle\int_0^L u(t,x,0)\,dx >0$ cf. \cite{CS0, CS}. For these waves therefore the (generalized) 
velocity potential $\v$ is not periodic in the $x$-variable. Thus, while $\v_x$ and $\v_y$ are both $x$-periodic 
with period $L$, the (generalized) potential $\v$ is not necessarily $L$-periodic in the $x$-variable. The 
above discussion shows that (7) is the necessary and sufficient condition for $\v$ to be 
$L$-periodic in the $x$-variable. For steady waves, the relation (7) means that the 
wave speed is defined as the mean velocity in the moving frame of reference in which the wave 
is stationary. Imposing (7) for the large-amplitude steady waves with vorticity studied in 
\cite{CS, CE, CSS, CS1, E, W} simply means that along the continuum of waves constructed in \cite{CS} 
the wave speed is not fixed {\it a priori}, but varies according to (7). 
As stated in the Introduction, throughout this paper we consider 
only flows satisfying (7) so that $\v$ is $L$-periodic in the $x$-variable. 

In terms of the functions $\v$ and $\psi$, we can recast the Euler equation (2) in the form
$$\nabla \left[ \varphi_t + \frac{1}{2} \,|\nabla \psi|^2 + P + \omega \psi + g y \right] = 0.$$
Thus at each instant $t \ge 0$,
\begin{equation}
\displaystyle\left[ \varphi_t + 
\frac{1}{2} \,|\nabla \psi|^2 + P + \omega \psi + g y \right]\quad\hbox{is constant throughout}\quad 
\Omega(t).
\end{equation} 
This is the generalization for flows of constant vorticity of 
Bernoulli's law \cite{S} for irrotational flows ($\omega=0$). In view of (3), we deduce that  
$$\displaystyle\left[ \v_t + 
\frac{1}{2} \,|\nabla \psi|^2 +\omega \psi + g \eta \right]\quad\hbox{is constant on the 
free surface}\quad S(t).$$
Since $\v$ is uniquely determined by (11) up to an arbitrary additive term that is solely 
time-dependent, we use this freedom to absorb into the definition of $\v$ a suitable 
time-dependent term so that
\begin{equation}
\v_t+\frac{1}{2} \,|\nabla \psi|^2 +\omega \psi + g \eta =0\quad\hbox{on}\quad S(t).
\end{equation} 

Let $\xi$ denote the evaluation of $\v$ at the free surface,
\begin{equation}
\xi(t,x)=\v(t,x,\eta(t,x)),\qquad t \ge 0, \ x \in [0,L].
\end{equation}
Given the constant vorticity $\omega$, the functions $\xi$ and $\eta$, taken to be smooth and 
$L$-periodic in the $X$-variable, completely determine the motion. Indeed, 
the function $x \mapsto \eta(t,x)$ fixes the fluid domain $\Omega(t)$, and $\xi(t,\cdot)$ is the 
appropriate boundary data for a linear elliptic problem of mixed type that determines 
$\v(t,\cdot,\cdot)$ at time $t \ge 0$. At any given time $t \ge 0$, fix $\eta$ and $\xi$, and let 
$\v$ be the unique solution of the boundary-value problem
\begin{equation}
\begin{cases}
\Delta\v=0\quad\hbox{in}\quad \Omega(t),\\
\v=\xi\quad\hbox{on}\quad S(t),\\
\v_y=0\quad\hbox{on}\quad y=0.
\end{cases}
\end{equation}
that is $L$-periodic in the $x$-variable. Knowing $\v$, we determine the 
velocity field $(u,v)$ from (11), the 
stream function $\psi$ from (10), and the pressure $P$ from Bernoulli's law (13).

The total energy of the wave motion in a period cell is given by
\begin{equation}
H=\iint_{\Omega(t)} \Big\{\frac{u^2+v^2}{2}+gy \Big\}\,dydx.
\end{equation}
In the above expression, the first term represents the kinetic energy (energy of motion), while 
the second term is the gravitational potential energy (energy of position). We now 
show that $H$ is completely determined by the functions $\xi$ and $\eta$. In order to 
do this, we introduce the {\it Hilbert transform}, $\T(\eta)$, defined as follows. 
At any given time $t \ge 0$, fix $\eta$ and $\xi$, and let 
$\v$ be the unique solution of the boundary-value problem (16). Since (8) and (11) ensure 
that $\displaystyle(\psi+\frac{\omega}{2}\, y^2)$ is the harmonic conjugate to $\v$, the Hilbert 
transform $\T(\eta)\,\xi$ of $\xi$ is given by 
\begin{equation}
\Big(\T(\eta)\,\xi\Big)(x)=\chi(t,x)+\frac{\omega}{2}\,\eta^2(t,x),\qquad x \in [0,L],
\end{equation}
where
\begin{equation}
\chi(t,x)=\psi(t,x,\eta(t,x)),\qquad t \ge 0, \ x \in [0,L].
\end{equation}
The equations (18)-(19) show that $\chi$ is completely determined by $\eta$ and $\xi$ via
\begin{equation}
\chi=\T(\eta)\,\xi-\frac{\omega}{2}\,\eta^2.
\end{equation}
Let
$$n=\frac{1}{\sqrt{1+\eta_x^2}}\,\begin{pmatrix}
-\eta_x \\
1
\end{pmatrix},$$
be the outward normal to the free surface $y=\eta(t,x)$. With $dl=\sqrt{1+\eta_x^2}\,dx$, 
using Green's identity, we get
\begin{eqnarray}
H &=& \frac{1}{2} \iint_{\Omega(t)} |\nabla\psi|^2dydx 
 + \frac{1}{2} \int_0^L g \,\eta^2\,dx \nonumber \\
&=& \frac{1}{2} \int_{S(t)} \psi\,[-\psi_x\eta_x+\psi_y]\,dx + \frac{\omega}{2} \iint_{\Omega(t)} \psi\,dydx
+ \frac{1}{2} \int_0^L g \,\eta^2\,dx 
\end{eqnarray}
in view of (10) and the fact that $\psi=0$ on $y=0$. Another application of Green's identity 
for the functions $\psi$ and $y^2$ yields
\begin{eqnarray*}
&&\iint_{\Omega(t)} \Big(-\,\omega\,y^2\,-\,2 \psi\Big)\,dydx = 
\iint_{\Omega(t)} \Big(y^2\,\Delta \psi\,- \psi\,\Delta(y^2)\Big)\,dydx \\
&&\quad = \int_{S(t)} \Big(y^2 
\,\frac{\partial \psi}{\partial n} -\frac{\partial y^2}{\partial n}\,\psi\Big)\,dl = 
\int_0^L (-\eta_x\psi_x+\psi_y)\,\eta^2\,dx -2\int_0^L \chi\,\eta\,dx
\end{eqnarray*}
if we take into account (10) and (19). Therefore
\begin{equation}
\iint_{\Omega(t)} \psi\,dydx = -\frac{\omega}{6} \int_0^L \eta^3\,dx  - 
\frac{1}{2} \int_0^L (-\eta_x\psi_x+\psi_y)\,\eta^2\,dx +\int_0^L \chi\,\eta\,dx.
\end{equation}
Taking into account (8), (11), we further transform the expression (21) for $H$ into 
$$ \frac{1}{2} \int_{S(t)} (\psi-\frac{\omega}{2}\,y^2)\,[\v_y\eta_x+\v_x-\omega\,y]\,dx
 + \frac{1}{2} \int_0^L (g -\frac{\omega^2}{6}\,\eta)\,\eta^2\,dx+\frac{\omega}{2}\int_0^L \chi\,\eta\,dx$$
in view of (22). Since
$$\xi_x=\v_x+\v_y\eta_x,\qquad x \in [0,L],$$
and using (18), we obtain
\begin{eqnarray}
H(\eta,\xi) &=& \frac{1}{2} \int_0^L \xi_x\,\cdot\T(\eta)\,\xi \,dx 
+\, \frac{1}{2} \int_0^L g \,\eta^2\,dx  \nonumber \\
&&\qquad -\, \frac{\omega}{2} \int_0^L  \xi_x\,\eta^2\,dx 
+ \frac{\omega^2}{6}\int_0^L \eta^3\,dx.
\end{eqnarray}
We now present the main result of this paper.\bigskip

{\bf Theorem 1} {\it The governing equations for periodic two-dimensional gravity water waves 
of constant vorticity $\omega$ are equivalent to the nearly-Hamiltonian system
\begin{equation}
\begin{cases}
\dot{\eta} = \displaystyle\frac{\delta H}{\delta \xi},  \vspace{0.3cm}\\ 
\dot{\xi} = -\,\displaystyle\frac{\delta H}{\delta \eta} - \omega \,\chi
\end{cases}
\end{equation}
with $y=\eta$ being the free surface, $\xi$ being the evaluation of the (generalized) velocity 
potential on the free surface, $H=H(\eta,\xi)$ being the total energy of the motion, and 
$$\chi=\T(\eta)\,\xi -\displaystyle\frac{\omega}{2}\,\eta^2$$
being the evaluation of the stream function on the free surface.}\bigskip  

{\bf Remark} For irrotational flows ($\omega=0$) the system $(24)$ is Hamiltonian: we recover 
Zakharov's result \cite{Z}.$\hfill\Box$\bigskip

Before proceeding with the proof of Theorem 1, let us review briefly the concept of an 
infinite-dimensional Hamiltonian system - see also \cite{CG, O}. An {\it infinite-dimensional 
Hamiltonian system} is a system of partial differential equations of the form
\begin{equation}
f_t=J\,\hbox{grad}\,H(f),
\end{equation}
where $f(t)$ describes a path in a Hilbert space endowed with an inner product $\langle \cdot, 
\cdot \rangle$, the {\it Hamiltonian function} $H: {\mathcal D} \to \R$ being defined on a dense 
subspace of the Hilbert space, and with $J$ being a skew-adjoint operator. The gradient in 
(25) is taken with respect to the inner product $\langle \cdot, \cdot \rangle$ on the 
Hilbert space. If the operator 
$J$ is invertible, this set-up defines a {\it symplectic structure} on the Hilbert space. The system 
(24) is nearly-Hamiltonian: one can view $\omega$ as a parameter measuring the deviation of 
(24) from a Hamiltonian structure of the form (25) with $f=\begin{pmatrix} \eta \\
\xi 
\end{pmatrix}$ on the Hilbert space $L^2[0,L] \times L^2[0,L]$, with operator 
$$J=\begin{pmatrix}
0 & 1\\
-1 & 0 
\end{pmatrix},$$     
the Hamiltonian function $H$ being given by (23) with $\eta$ and $\xi$ in the dense subspace 
${\mathcal D} \subset L^2[0,L]$ of smooth $L$-periodic functions.  \bigskip

{\it Proof of Theorem 1.} Let us first compute the $\xi$-gradient of 
$H$, $\displaystyle\frac{\delta H}{\delta\xi}$. We vary 
$\xi$ and keep $\eta$ fixed. If $\theta$ is a harmonic function in $\Omega(t)$, $L$-periodic in the 
$x$-variable and with $\theta_y=0$ on $y=0$, let $\theta_0$ be the evaluation of $\theta$ 
on $S(t)$. If $\Psi$ is the harmonic conjugate 
of $\theta$ with $\Psi=0$ on $y=0$, from (18) we get
$$\T(\eta)\,[\xi]=\chi+\frac{\omega}{2}\,\eta^2,\qquad \T(\eta)\,[\xi+\varepsilon\,\theta_0]=\chi+\varepsilon\,\zeta+\frac{\omega}{2}\,\eta^2,$$
denoting by $\zeta$ the evaluation of $\Psi$ on $S(t)$. Therefore
\begin{eqnarray}
\langle \frac{\delta H}{\delta\xi},\, \theta_0 \rangle &=& \lim_{\varepsilon \to 0} 
\frac{H(\eta,\xi+\varepsilon \theta_0)-H(\eta,\xi)}{\varepsilon} \nonumber \\
&=& -\frac{1}{2} \int_0^L \Big\{ \zeta_x\,\xi + (\chi_x-\omega\,\eta\,\eta_x)\,\theta_0\,)\,dx 
\end{eqnarray}
in view of (23) and using the periodicity. Since
$$\theta_x=\Psi_y,\qquad \theta_y=-\Psi_x\quad\hbox{throughout}\quad \Omega(t),$$
we obtain
\begin{eqnarray}
&&\int_0^L \zeta_x\,\xi\,dx = \int_{S(t)} (\Psi_x+\Psi_y\eta_x)\,\v\,dx =
 \int_{S(t)} (-\theta_y+\theta_x\,\eta_x)\,\v\,dx \nonumber \\
&&\quad = -\int_{S(t)} \frac{\partial \theta}{\partial n}\,\v\,dl 
= -\int_{S(t)} \theta\,\frac{\partial\v}{\partial n}\,dl  =\int_{S(t)} \theta_0 \,(\v_x\eta_x-\v_y)\,dx 
\end{eqnarray}
using in the next to last step Green's identity for the harmonic functions $\v$ and $\theta$, which 
satisfy $\theta_y=\v_y=0$ on $y=0$. From (26)-(27) we infer that
$$\langle \frac{\delta H}{\delta\xi},\, \theta_0 \rangle = -\frac{1}{2} \int_{S(t)} 
( \v_x\eta_x-\v_y+\chi_x-\omega\,\eta\eta_x )\,\theta_0\,dx= \int_0^L (v-u\,\eta_x)\,\theta_0\,dx,$$
if we take into account (8), (11) and (19). Thus
$$\frac{\delta H}{\delta\xi}=v-u\,\eta_x.$$
Since $\dot{\eta}=\eta_t$, we can recast (4) as
$$\dot{\eta}=\frac{\delta H}{\delta\xi}.$$

To recover the remaining part of the system (24), we have to compute 
$\displaystyle\frac{\delta H}{\delta \eta}$. To do this, instead of working with the 
expression (23), it is simpler to compute the variation in (17). This calculation is not 
straigthforward due to the fact that there is an implicit nonlinear dependence of $\v$ 
upon $\eta$. As an example of the intricacies of the calculation, notice that, 
contrary to a possible first impression, the Hilbert transform 
$\T(\eta)$ is not skew-adjoint if the surface $y=\eta(t,x)$ is not flat (see \cite{C, CG} 
for a discussion). 

Let
\begin{equation}
\begin{cases}
\xi_1(t,x)=\v_x(t,x,\eta(t,x)),\\
\xi_2(t,y)=\v_y(t,x,\eta(t,x)),
\end{cases}
\end{equation}
be the evaluations of the partial derivatives of the potential $\v$ on the free surface $y=\eta(t,x)$. 
Notice that $\xi_1$ is not to be confused with the function $\xi_x$ since 
$$\xi_x=\xi_1+\eta_x\xi_2.$$

In terms of the potential $\v$, we can rewrite (17) as 
\begin{eqnarray}
H(\eta,\v)=  - \i\j \, y \, \varphi_x \,dydx  &+& 
\frac{1}{2} \i\j \omega^2 y^2 \,dydx \nonumber \\ 
&+&   \frac{1}{2} \i\j |\nabla \v|^2 \,dy  +  \frac{1}{2} \i\j\, g \, \eta^2 \, dx.
\end{eqnarray}
We now vary the function $\eta$ describing the free surface by $\d\eta$, keeping $\xi$ fixed. 
Since $\v$ and $(\psi+\displaystyle{\omega \over 2}\,y)$ are harmonic conjugate functions, 
by analytic continuation \cite{B} the function $\v$ has a harmonic extension across the boundary. 
Varying the domain $\Omega(t)$ to $\Omega_\varepsilon(t)$ by keeping $\xi(t,\cdot)$ fixed means that 
we solve instead of (16) the problem
$$\begin{cases}
\Delta\v_\varepsilon=0\quad\hbox{in}\quad \Omega_\varepsilon(t),\\
\v_\varepsilon=\xi\quad\hbox{on}\quad S_\varepsilon(t),\\
\partial_y\,\v_\varepsilon=0\quad\hbox{on}\quad y=0,
\end{cases}$$
where
\begin{eqnarray*}
\Omega_\varepsilon(t) &=& \{(x,y) \in \R^2:\ 0<x<L,\ 0<y<\eta(t,x)+\varepsilon\,(\delta\eta)(t,x)\},\\
S_\varepsilon(t) &=& \{(x,y) \in \R^2:\ 0<x<L,\ y=\eta(t,x)+\varepsilon\,(\delta\eta)(t,x)\}.
\end{eqnarray*}
Since
$$\v_\varepsilon(t,x,\eta+\varepsilon\,\delta\eta)=\v(t,x,\eta)=\xi(t,x)$$
we deduce that
\begin{eqnarray*}
{\v_\varepsilon(t,x,\eta) -\v(t,x,\eta) \over \varepsilon} &=& {\v_\varepsilon(t,x,\eta)-\v_\varepsilon(t,x,\eta+
\varepsilon\,\delta\eta) \over \varepsilon} \\
&=& -\,\delta\eta\,\cdot\, \partial_y\,\v_\varepsilon(t,x,\eta) + O(\varepsilon) \\
&\to& - \,\delta\eta\,\cdot \v_y(t,x,\eta)\quad\hbox{as}\quad \varepsilon \to 0.
\end{eqnarray*}
Therefore, if $[\d\v]^\#$ is the evaluation of the variation $\d\v$ of $\v$ on 
$y=\eta(t,x)$, we showed that
\begin{equation}
[\d\v]^\#=-\,\xi_2\,\d\eta.
\end{equation}
Other than a variation $\delta\v$ of $\v$, we have the variation
\begin{eqnarray}
\d H & =  & - \i  \j \omega \, y \, \delta \v_x \, dydx
- \i \omega \, \eta \, \xi_1 \, \delta \eta \, dx
+ \frac{1}{2} \i \omega^2 \eta^2 \, \delta \eta \, dx
\nonumber \\ 
&&\quad + \i  \j (\nabla \varphi) \cdot \nabla \delta \varphi \,dydx  +
\frac{1}{2} \i (\xi_1^2+\xi_2^2) \, \delta \eta \, dx + \i g \, \eta \, \delta \eta \, dx 
\end{eqnarray}
of $H$, since $\d\nabla\v=\nabla\d\v$. Using the formula
$$\partial_x \, \j F(x, y) \, dy \, =  \j F_x(x, y) \, dy \, + \, F[x, \eta] \,\, \eta_x \, , $$
the first term on the right-hand side of (31) can be rewritten as
\begin{eqnarray}
\i  \j \omega \, y \, \delta \varphi_x \, dydx &=&  \i \left[ \j \omega \, y \, \delta \varphi \, dy
\right]_x \, dx  - \i \omega \, \eta \, [\d\v]^\# \, \eta_x \,dx \nonumber \\
&=& -\,\i \omega\,\eta\,[\d\v]^\#\,\eta_x\,dx=\omega\int_0^L \eta\eta_x\xi_2\d\eta\,dx
\end{eqnarray}
by (30) and periodicity. Combining (31)-(32), we obtain
\begin{eqnarray}
\delta H  &=&  \i \Big(- \omega \, \eta \, \eta_x \, \xi_2  \, - \, \omega \, \eta \, \xi_1 \, + \,
\frac{1}{2} \,\omega^2\, \eta^2
+ \, \, \frac{1}{2} (\xi_1^2 +\xi_2^2) \, + \, g \, \eta \Big)\, \delta \eta \, dx \nonumber \\
&&\qquad + \i  \j (\nabla \varphi) \cdot \nabla \delta \varphi \,\, dydx \, .
\end{eqnarray}
Since $\v$ is harmonic in $\Omega(t)$, applying Green's identity to the pair of functions 
$\v$ and $\d\v$, we infer that the last integral in the above expression equals
$$\int_{\partial\Omega(t)} \d\v\,\frac{\partial\v}{\partial n}\,dl=-\,
\int_0^L \xi_2\,[\xi_2-\xi_1\,\eta_x]\,\d\eta\,dx$$
in view of (5) and (11), respectively (28) and (30). Therefore (33) becomes
$$\d H=\int_0^L \Big( -\omega\,\eta\eta_x\xi_2-\omega\,\eta\xi_1+
\frac{1}{2}\,\omega^2\eta^2+\frac{\xi_1^2+\xi_2^2}{2}+g\,\eta-\xi_2^2
+\xi_1\xi_2\eta_x\Big)\,\d\eta\,dx.$$
Thus
\begin{equation}
\frac{\d H}{\d\eta}=-\omega\,\eta\eta_x\xi_2-\omega\,\eta\xi_1+
\frac{1}{2}\,\omega^2\eta^2+\frac{\xi_1^2+\xi_2^2}{2}+g\,\eta-\xi_2^2
+\xi_1\xi_2\eta_x.
\end{equation}
Combining (14) with (8), (11), (19), and (28), we obtain
$$\v_t+\frac{\xi_1^2+\xi_2^2}{2}-\omega\eta\xi_1+\frac{1}{2}\,\omega^2\eta^2
+\omega\,\chi+g\eta=0 \quad\hbox{on}\quad y=\eta(t,x),$$
so that (34) becomes
\begin{equation}
\frac{\d H}{\d \eta}=-\v_t-\omega\eta\eta_x\xi_2-\xi_2^2+\eta_x\xi_1\xi_2-\omega\chi.
\end{equation}
Differentiating (15) with respect to $t$, we get 
$$\xi_t=\v_t+\xi_2\eta_t\quad\hbox{on}\quad y=\eta(t,x),$$
in view of (28). Furthermore, (4), (11) and (28) yield
$$\xi_2=\eta_t+(\xi_1-\omega\eta)\eta_x\quad\hbox{on}\quad y=\eta(t,x).$$
From the previous two relations and (35) we infer
$$\frac{\d H}{\d\eta}=-\xi_t-\omega\chi,$$
which completes the proof.$\hfill\Box$

\section{Hamiltonian structure for steady waves}

In this section we consider the case of steady water waves with constant vorticity.  
If $c \neq 0$ is the speed of the wave, (4) becomes
$$-\psi_x=-c\eta_x+\psi_y\eta_x\quad\hbox{on}\quad y=\eta(x-ct),$$
if we use (8). Thus $\partial_x\,\Big[\psi\Big(x-ct,\eta(x-ct)\Big)-\,c\,\eta(x-ct)\Big]=0$ so 
that in this setting the function $[\psi-cy]$ is constant on the free surface. On the other 
hand, using (1), we deduce that
\begin{eqnarray*}
\partial_x \,\Big[ \int_0^\eta (u-c)\,dy\Big] &=& \Big(u(x-ct,\eta(x-ct))\,-c\Big)\,\eta_x(x-ct) 
+\int_0^\eta u_x\,dy \\
&=& \Big(u(x-ct,\eta(x-ct))\,-c\Big)\,\eta_x(x-ct) 
-\int_0^\eta v_y\,dy \\
&=& \Big(u(x-ct,\eta(x-ct))\,-c\Big)\,\eta_x(x-ct) -v(x-ct,\eta(x-ct))
\end{eqnarray*}
has to equal zero by (1). Since $\psi_y=u$ throughout the fluid and $\psi=0$ on the flat bed, we have
$$\int_0^\eta  (u-c)\,dy=(\psi-cy)\Big|_{y=\eta(x-ct)}.$$
The constant value $k$ of the expression on the above left-hand side is the {\it relative 
mass flux} of the flow \cite{CS}. Since field evidence and laboratory measurements indicate that 
for steady water waves that are not near the breaking state the relation $u<c$ 
holds throughout the fluid \cite{L}, we have that $k<0$. These considerations show in view of 
Theorem 1 that the governing equations for steady water waves with constant vorticity $\omega$ 
and wave speed $c$ are equivalent to the nearly-Hamiltonian system
\begin{equation}
\begin{cases}
\dot{\eta} = \displaystyle\frac{\delta H}{\delta \xi},  \vspace{0.3cm}\\ 
\dot{\xi} = -\,\displaystyle\frac{\delta H}{\delta \eta} - \omega \,(k+c\eta)
\end{cases}
\end{equation}
where $k<0$ is the relative mass flux of the flow. From (36) we readily obtain the Hamiltonian 
formulation for steady water waves with constant vorticity.\bigskip

{\bf Theorem 2} {\it The governing equations for steady $L$-periodic two-dimensional 
gravity water waves with constant vorticity $\omega$ are equivalent to the 
Hamiltonian system
\begin{equation}
\begin{cases}
\dot{\eta} = \displaystyle\frac{\delta \hat{H}}{\delta \xi},  \vspace{0.3cm}\\ 
\dot{\xi} = -\,\displaystyle\frac{\delta \hat{H}}{\delta \eta}\, .
\end{cases}
\end{equation}
Here $c \neq 0$ is the speed of the wave, $y=\eta$ is the free surface above the flat bed $y=0$, 
$\xi$ is the evaluation of the (generalized) velocity potential on the free surface, and
$$\hat{H}=H +\omega k \int_0^L \eta\,dx+\frac{c\omega}{2}\int_0^L \eta^2\,dx,$$
where $H=H(\eta,\xi)$ is the total energy of the motion.}\bigskip  

{\bf Remark} For the Hamiltonian system (37) we have 
$\T(\eta)\xi-c\eta-\displaystyle\frac{\omega}{2}\,\eta^2=k$ 
in view of (20), since $\xi=k+c\eta$.$\hfill\Box$\bigskip

\vskip.5cm
\n
{\it Acknowledgement} It is a pleasure to thank David Kaup for very useful discussions. 
We also thank both referees for constructive comments and suggestions. 
The financial support of Science Foundation Ireland (AC and EMP, grant 04/BRG/M0042)
and of the Irish Research Council for Science, Engineering and Technology (RII) 
is gratefully acknowledged.

\providecommand{\href}[2]{#2}


\begin{thebibliography}{10}

\bibitem{AT}
C.~J.~Amick and J.~F.~Toland, On periodic water-waves and their convergence to solitary 
waves in the long-wave limit, {\it Philos. Trans. Roy. Soc. London Ser. A} {\bf 303} (1981), 
633--669.

\bibitem{BO}
T.~B.~Benjamin and P.~J.~Olver,  Hamiltonian structure, symmetries and conservation laws for water waves, 
{\it J. Fluid Mech.} {\bf 125} (1982), 137--185.

\bibitem{B}
R.~P.~Boas, {\it Invitation to Complex Analysis}, McGraw-Hill, Inc., New York, 1987.

\bibitem{CDG}
J.~Cantarella, D.~DeTurck, and H.~Gluck, Vector calculus and the topology of 
domains in 3-space, {\it Amer. Math. Monthly} {\bf 109} (2002), 409--442.

\bibitem{Co0}
A.~Constantin, A Hamiltonian formulation for free surface water waves with non-vanishing 
vorticity, {\it J. Nonl. Math. Phys.} {\bf 12} (2005), 202--211.

\bibitem{Co}
A.~Constantin, Wave-current interactions, in {\it EQUADIFF 2003}, pp. 207--212, World Sci. 
Publ., Hackensack, NJ, 2005.

\bibitem{Co2}
A.~Constantin, The trajectories of particles in Stokes waves, {\it Inv. Math.}, in print.

\bibitem{CE}
A.~Constantin and J. Escher, Symmetry of steady periodic surface water waves with vorticity, 
{\it J. Fluid Mech.} {\bf 498} (2004), 171--181.

\bibitem{CS0} 
A.~Constantin and W.~Strauss, Exact periodic traveling water waves with vorticity, {\it C. R. 
Math. Acad. Sci. Paris} {\bf 335} (2002), 797--800.

\bibitem{CS}
A.~Constantin and W.~Strauss, Exact steady periodic water waves with vorticity, 
{\it Comm. Pure Appl. Math.} {\bf 57} (2004), 481--527.

\bibitem{CSS}
A.~Constantin, D.~H.~Sattinger, and W.~Strauss, Variational formulations for steady water 
waves with vorticity, {\it J. Fluid Mech.} {\bf 548} (2006), 151--163

\bibitem{CS1}
A.~Constantin and W. Strauss, Stability properties of steady water waves with vorticity, 
{\it Comm. Pure Appl. Math}, in print.

\bibitem{C}
W.~Craig, Water waves, Hamiltonian systems and Cauchy integrals, in {\it Microlocal Analysis 
and Nonlinear Waves} (Minneapolis, MN, 1988--1989), pp. 37--45, IMA Vol. Math. Appl., 
30, Springer, New York, 1991.

\bibitem{CG}
W.~Craig and M.~D.~Groves, Hamiltonian long-wave approximations to the water-wave problem, 
{\it Wave Motion} {\bf 19} (1994), 367--389.

\bibitem{DP}
A.~F.~T.~DaSilva and D.~H.~Peregrine, Steep, steady surface waves on water of finite depth 
with constant vorticity, {\it J. Fluid Mech.} {\bf 195} (1988), 281--302.

\bibitem{E}
M.~Ehrnstr\"om, Uniqueness for steady water waves with vorticity, {\it Int. Math. Res. Not.} 
{\bf 60} (2005), 3721--3726.

\bibitem{GT}
M.~D.~Groves and J.~F.~Toland, On variational formulations for steady water waves, 
{\it Arch. Rat. Mech. Anal.} {\bf 137} (1997), 203--226.

\bibitem{J}
R.~S.~Johnson, {\it A Modern Introduction to the Mathematical Theory of Water Waves}, 
Cambridge University Press, Cambridge, 1997.

\bibitem{KS}
B.~Kolev and D.~H.~Sattinger, Variational principles for stationary waves, {\it SIAM 
J. Math. Anal.}, in print.

\bibitem{L}
J.~Lighthill, {\it Waves in Fluids}, Cambridge University Press, Cambridge, 1978.

\bibitem{O}
P.~J.~Olver, {\it Applications of Lie Groups to Differential Equations}, Springer Verlag, 
New York, 1986.

\bibitem{S}
J.~J.~Stoker, {\it Water Waves. The Mathematical Theory with Applications}, 
Interscience Publ., New York, 1957. 

\bibitem{St}
G.~Stokes, On the theory of oscillatory waves, {\it Trans. Cambridge Phil. Soc.} {\bf 8} 
(1847), 441--455.

\bibitem{SCJ}
C.~Swan, I.~Cummins, and R. James, An experimental study of two-dimensional 
surface water waves propagating on depth-varying currents, {\it J. Fluid Mech.} 
{\bf 428} (2001), 273--304.

\bibitem{Th}
G.~Thomas, Wave-current interactions: an experimental and numerical study, 
{\it J. Fluid Mech.} {\bf 216} (1990), 505--536. 

\bibitem{T}
J.~F.~Toland, Stokes waves, {\it Topol. Meth. Nonl. Anal.} {\bf 7} (1996), 1--48.

\bibitem{W}
E.~Wahlen, A note on steady gravity waves with vorticity, {\it Int. Math. Res. Not.} 
{\bf 7} (2005), 389--396.

\bibitem{W0}
F.~W.~Warner, {\it Foundations of Differentiable Manifolds and Lie Groups}, Springer Verlag, 
New York, 1983.

\bibitem{Z} V.~E.~Zakharov, Stability of periodic waves of finite amplitude on the surface 
of a deep fluid, {\it J. Appl. Mech. Tech. Phys.} {\bf 2} (1968), 190--194.

\end{thebibliography}
\end{document}